\documentclass[aps, prx,twocolumn,longbibliography,superscriptaddress,showpacs,amsmath,amssymb,floatfix]{revtex4}
\usepackage[latin9]{inputenc}
\usepackage{amssymb}
\usepackage{graphicx}
\usepackage{amsmath}
\usepackage{color}
\usepackage{mathrsfs}
\usepackage{float}
\usepackage{indentfirst}
\usepackage{mathrsfs}
\usepackage{float}
\usepackage{indentfirst}
\usepackage{textcomp}
\usepackage{comment}
\usepackage{mathtools}
\usepackage{natbib,hyperref}
\usepackage{soul}

\begin{document}

\title{On the Magnetization of the $120^\circ$ order of the Spin-1/2 Triangular Lattice Heisenberg Model: a DMRG revisit}

\author{Jiale Huang}
\affiliation{Key Laboratory of Artificial Structures and Quantum Control (Ministry of Education),  School of Physics and Astronomy, Shanghai Jiao Tong University, Shanghai 200240, China}

\author{Xiangjian Qian}
\affiliation{Key Laboratory of Artificial Structures and Quantum Control (Ministry of Education),  School of Physics and Astronomy, Shanghai Jiao Tong University, Shanghai 200240, China}

\author{Mingpu Qin} \thanks{qinmingpu@sjtu.edu.cn}
\affiliation{Key Laboratory of Artificial Structures and Quantum Control (Ministry of Education),  School of Physics and Astronomy, Shanghai Jiao Tong University, Shanghai 200240, China}
\affiliation{Hefei National Laboratory, Hefei 230088, China}

\begin{abstract}

We revisit the issue about the magnetization of the $120^\circ$ order in the spin-1/2 triangular lattice Heisenberg model (TLHM) with Density Matrix Renormalization Group (DMRG). The accurate determination of the magnetization of this model is challenging for numerical methods and its value exhibits substantial disparities across various methods. We perform a large-scale DMRG calculation of this model by employing bond dimension as large as $D = 24000$ and by studying the system with width as large as $L_\mathrm{y} = 12$.
With careful extrapolation with truncation error and suitable finite size scaling, we give a conservative estimation of the magnetization as $M_0 = 0.208(8)$. The ground state energy per site we obtain is $E_g = -0.5503(8)$. Our results provide valuable benchmark values for the development of new methods in the future.

\end{abstract}

\maketitle

\section{Introduction}
\label{sec:I}

The triangular lattice Heisenberg model (TLHM) with $S = 1/2$ is the simplest model to study the effect of frustration
in quantum magnetism. This model's ground state has remained as a topic of intense interest. From the 1970s, there was speculation 
that this model might host a resonating valence bond (RVB) quantum spin liquid (QSL), a state characterized by long-range
quantum entanglement and a conspicuous absence of magnetic order \cite{ANDERSON1973153,RVB1987,QSL1988,QSL1993}. However,
subsequent research has definitively established that the ground state of TLHM possesses a three-sublattice $120^\circ$ magnetic order \cite{order1988,order1991,order1992,SE1993,GFMC}. 
This revelation has profoundly influenced our understanding of this model and its fundamental characteristics.

Though consensus has been reached that the ground state of TLHM has the three-sublattice $120^\circ$ order, certain aspects of this state's properties remain elusive, particularly concerning the amplitude of the magnetization, which has been a subject of significant debate. Over the past several decades, various analytical and numerical methods have been employed to study the TLHM \cite{SWT1989,SWT1992,SE1993,QMC1997,GFMC,ED2004,VMC2006,SE2006,FN2006,White2007,VMC2009,VMC2014,CC2014,CC2015,SB2015,CC2016,SB+1/N,PESS2022,order2014,BDMC2013,SWT2009,Aorder1993}. Despite the effectiveness of these methods in addressing other aspects of this model \cite{Phase1992,Phase1993,Phase1995,SWTPhase1993,DMRG2016,CC2016}, they have yielded magnetization results for TLHM which exhibit considerable variations, with the maximum value almost twice of the minimum value. Nowadays, with the development of very accurate methods \cite{PhysRevX.4.011025,RevModPhys.93.045003,TNN2023,PhysRevLett.129.227201,Qian_2023} and the increase of computational power, it is possible to determine the precise value of the magnetization of TLHM by reaching consensus among different state-of-art methods.

In this work, we revisit this issue by utilizing the Density Matrix Renormalization Group (DMRG) \cite{DMRG1992,DMRG1993,SCHOLLWOCK201196} method. Previous calculation with DMRG gives a value of magnetization as $M_0 = 0.205(15)$ \cite{White2007}. Thanks to advancements in computational capabilities and improvement in algorithm details, we can now reach unprecedented bond dimension as large as $D = 24000$ in DMRG which gives us more accurate results and enables us to handle wider systems (with width $12$). 

With careful extrapolation with truncation error and following the similar procedure in Ref \cite{White2007} by varying the aspect ratio of the studied system to reduce the finite size effect,  we give a conservative estimation of the magnetization as $M_0 = 0.208(8)$. Notably, this value is in reasonable agreement with some of the results in the literature \cite{SE2006,SB2015,CC2015,CC2016}.
Furthermore, we have determined the ground state energy per site to be $E_g = -0.5503(8)$. Remarkably, this result exhibits a high degree of consistency with values from a range of numerical methods \cite{DMRG2016,CC2016,SE2006,PESS2022}.

The rest of the paper is organized as follows. In Sec.~\ref{sec:II}, we introduce the model and the methods used in this work. In Sec.~\ref{sec:III}, we present the results from DMRG, including the magnetization $M_0$ and ground state energy $E_g$. Finally, we conclude this work in Sec.~\ref{sec:IV}.

\section{Model and Methods}
\label{sec:II}

\subsection{Model}
The Hamiltonian of the spin-1/2 TLHM is given by

\begin{equation}
	H = J\sum_{\langle i,j \rangle} \hat{S}_i \cdot\hat{S}_j
	\label{eq:ham}
\end{equation}
with $\langle i,j\rangle$ denoting the nearest neighboring sites. $\hat{S}_i = (S^x_i,S^y_i,S^z_i)$ is the spin-1/2 operator at site $i$, and $J > 0$ is the antiferromagnetic exchange coupling which is set $J = 1$.

\begin{figure}[t]
	\includegraphics[width=85mm]{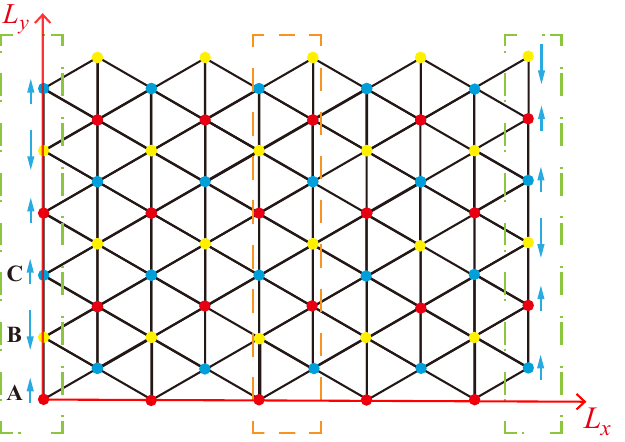}
	\caption{The illustration of a $6\times10$ triangular lattice. Periodic (open) boundary conditions are imposed in the y(x) directions. The red, yellow, and blue dots represent the three sublattices A, B, and C. The green dashed rectangular denotes the open edges, where the magnetic pinning fields are applied in the $z$ direction. The blue arrow represents the direction and value of the pinning magnetic fields with strength $0.25$, $-0.5$, $0.25$ for the A, B, and C sublattices respectively. The lattices in the middle column (orange dashed rectangular) are used to calculate $M$ (defined in Eq.~\ref{eq:m}).}
	\label{Lattice}
\end{figure}

\begin{figure*}[t]
	\centering
	\includegraphics[height=45mm]{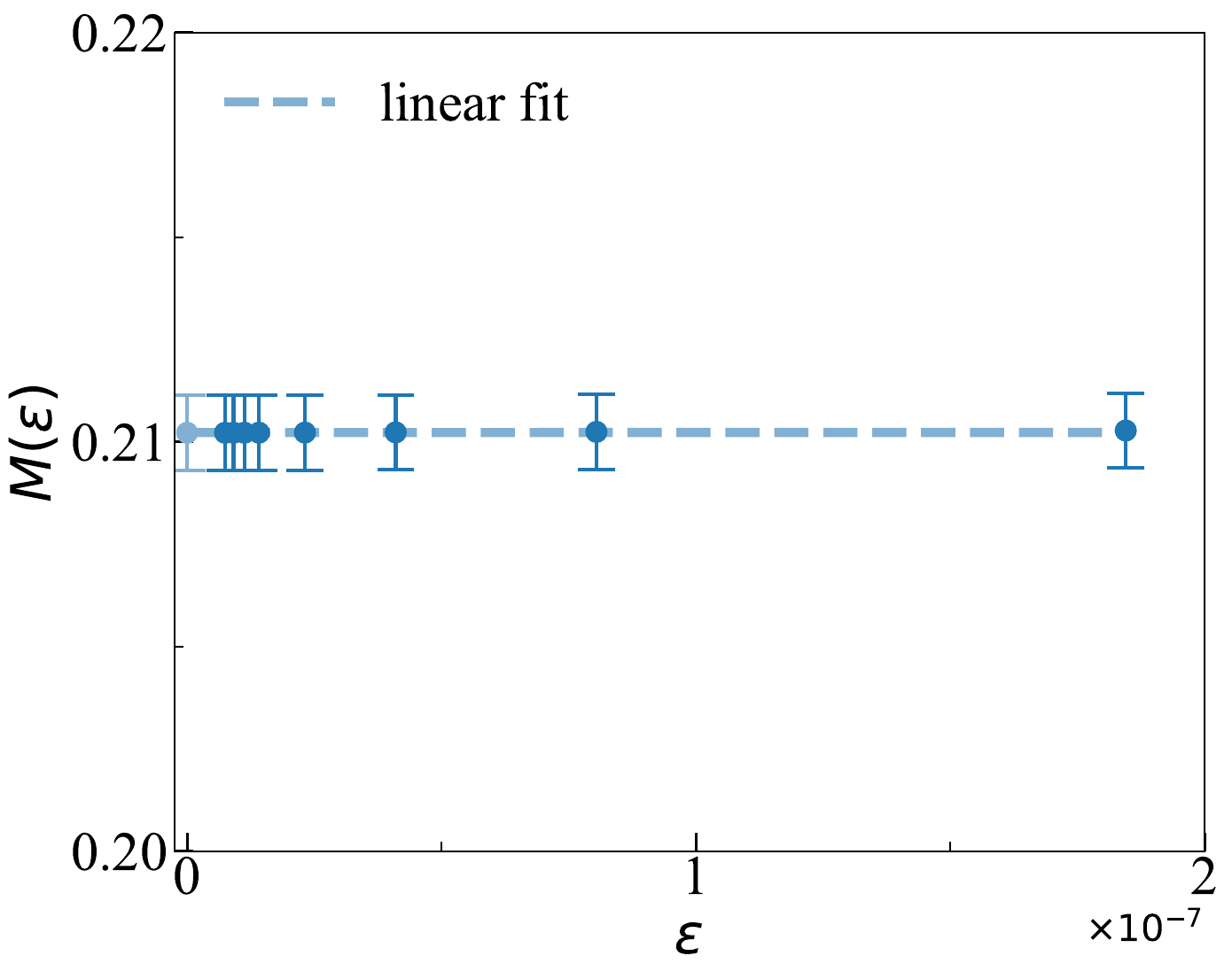}
	\includegraphics[height=45mm]{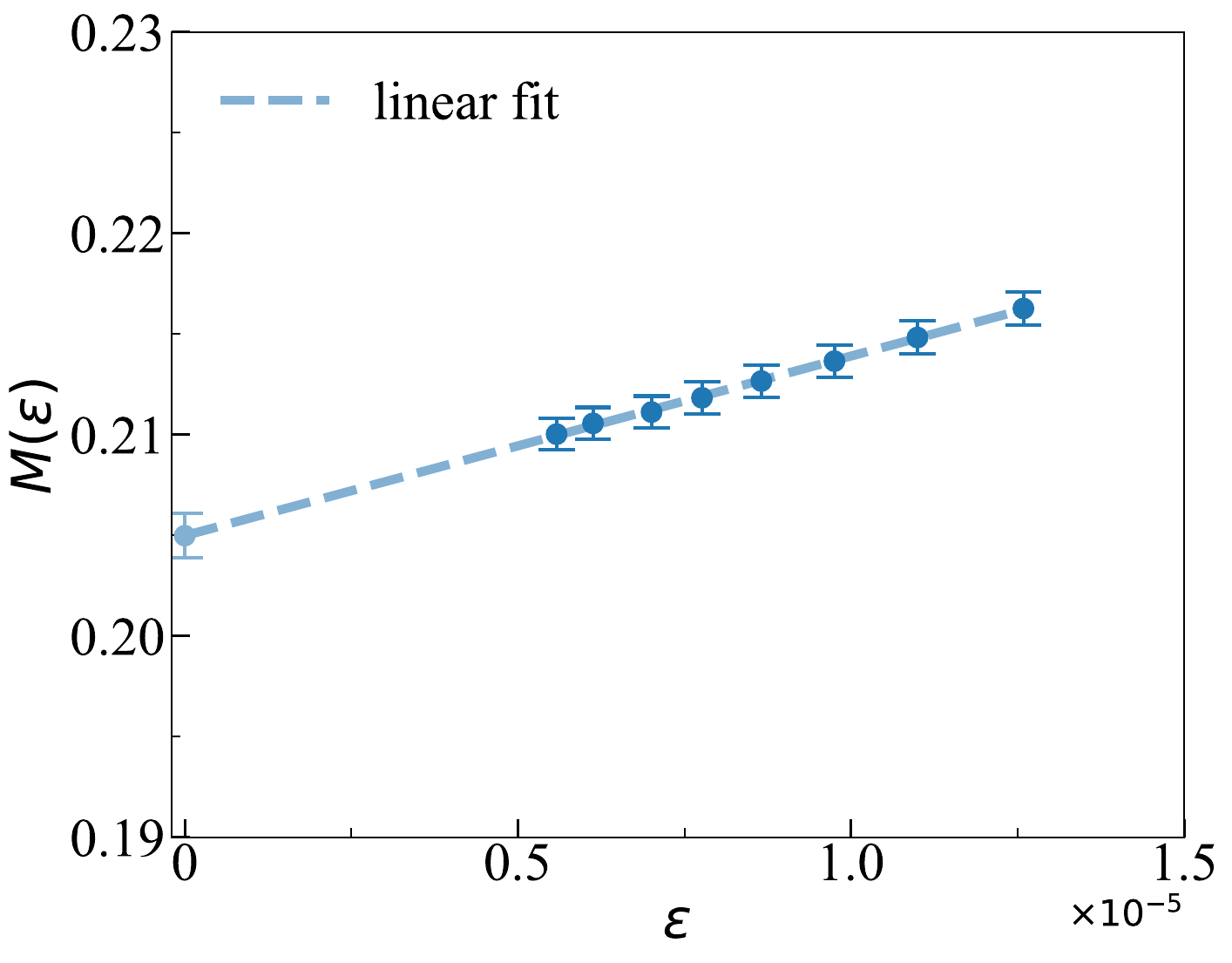}
	\includegraphics[height=45mm]{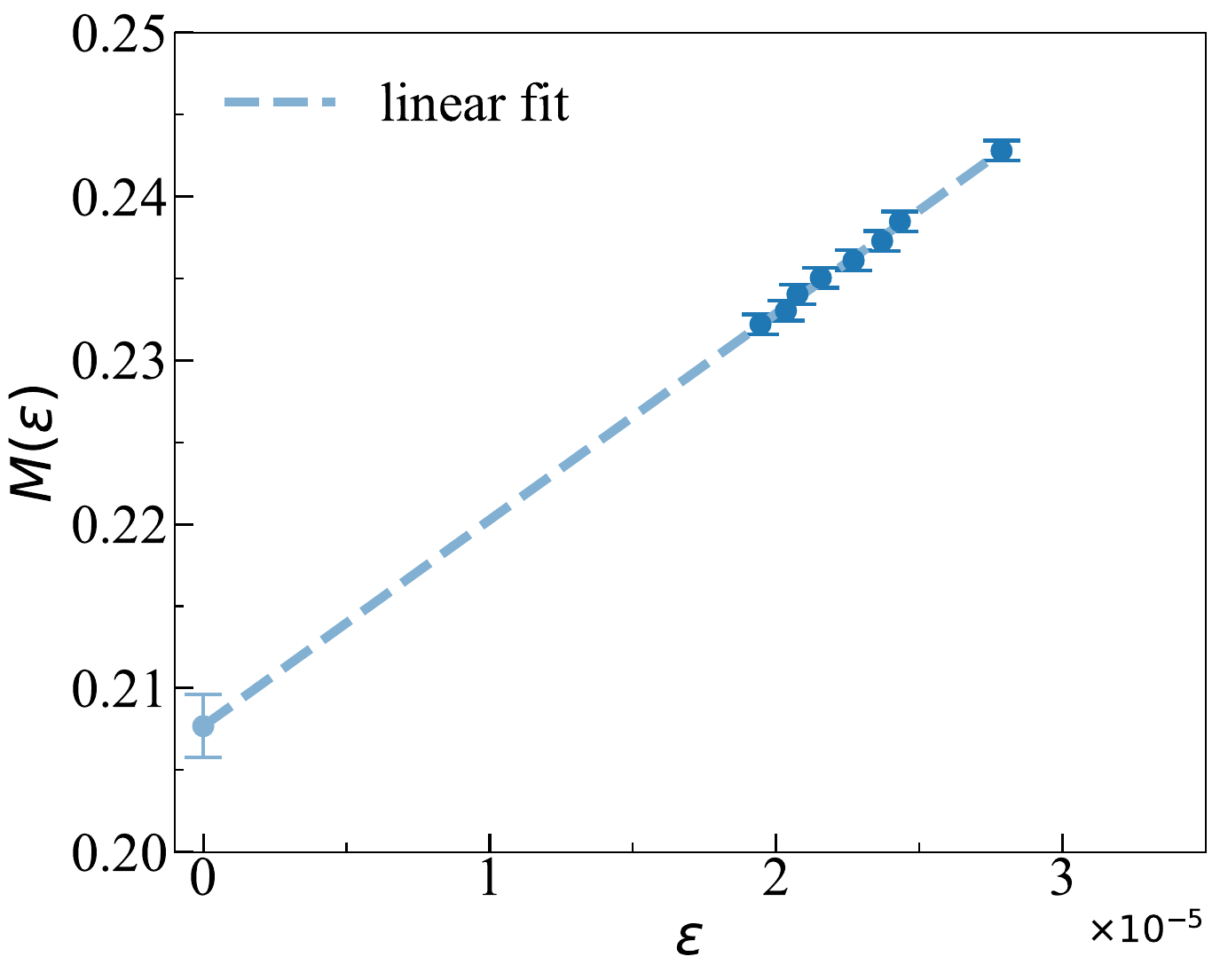}
	\caption{The variation of the averaged magnetization $M$ (defined in Eq.~(\ref{eq:m})) in the center column as a function of truncation error $\varepsilon$ in DMRG calculation. The aspect ratio is $\alpha=4/3$ and we show results for $L_\mathrm{y}=6, 9$, and $12$. The dashed lines represent linear fits. The extrapolated to $\varepsilon = 0$ results for $L_\mathrm{y} =6, 9$, and $12$ are $M =0.210(1), 0.205(1)$, and $0.208(2)$ respectively.}
	\label{Fit_4/3}
\end{figure*}

\subsection{Method}

DMRG is now arguably the workhorse for one-dimensional quantum systems. It doesn't suffer the notorious sign problem \cite{NegativeSign1990} but the limited entanglement encoded in the underlying wavefunction ansatz jeopardizes its accuracy in the study of real two-dimensional (2D) systems. When studying a 2D system, the required bond dimension grows exponentially with the width $L_\mathrm{y}$ of the system if we want to maintain the accuracy. However, with the increase of computational power and advances in algorithm realization, we can now handle a relatively wide ($12$ in this work) cylindrical system with decent accuracy by pushing the bond dimension to large values.

The underlying wavefunction ansatz of DMRG is the so called matrix product state (MPS) \cite{MPS} in the following form,
\begin{equation}
	|\Psi\rangle = \sum_{\{s_i\}}  A^{s_1} A^{s_2} \cdots A^{s_{N}} |s_1 s_2 \cdots s_{N}\rangle
	\label{eq:mps}
\end{equation}
where $N = L_\mathrm{x} \times L_\mathrm{y}$ is the number of total sites. $A^{s_i}$ is the local tensor at site $i$, and $s_i$ is the physical index. The bond dimension $D$ of the local tensor $A$ is a key parameter in DMRG calculations which determines the accuracy of the calculation. In this study, we push $D$ to as large as $24000$, with which we can reach the convergence regime of the DMRG calculation. Then we can eliminate the residual finite bond dimension error with an extrapolation with truncation error.

In Fig.~\ref{Lattice} we give an illustration of the triangular lattice. We study systems with cylindrical geometry by adopting open boundary conditions (OBC) in the $L_\mathrm{x}$ direction and periodic boundary conditions (PBC) in the $L_\mathrm{y}$ direction, which is easier for DMRG calculation. 

To calculate the magnetization in the system, we follow the strategy in Ref \cite{White2007} by applying magnetic pinning fields in the z-direction at the boundary of the open edges. By explicitly breaking the SU(2) symmetry, this scheme enables us to directly calculate the local magnetization instead of the more demanding spin-spin correlation function. Moreover, the entanglement in the ground state for the system with edge pinning fields is smaller than the SU(2)-symmetric one, making the convergence with bond dimension easier for DMRG calculation. 

It is known that the ground state of TLHM hosts the $120^\circ$ three-sublattice order, so to accommodate this order, the strengths of the pinning fields in the z-direction at the open edges are set as $0.25, -0.5, 0.25$ with ratio $1: -2: 1$. We only consider the magnetization in the middle column of the system (see the orange rectangular in Fig.~\ref{Lattice}) and calculate the weighted averaged value of the three sublattices as the magnetization for a given system as follows,
\begin{equation}
M = (\langle S^z_B \rangle_{\mathrm{avg}} - 2\langle S^z_A \rangle_{\mathrm{avg}} - 2 \langle S^z_C \rangle_{\mathrm{avg}})/3
\label{eq:m}
\end{equation}
where $\langle A,B,C \rangle$ means the three sublattices located in the central column of the system.
With the given pinning fields, we expect that the magnetizations in the z-direction for the three sublattices have a ratio $-1: 2: -1$. But for finite sizes, the actual ratio deviates from this ideal value. This effect is reflected in the error bar of the averaged value in Eq.~(\ref{eq:m}).

To obtain the value of magnetization in the two-dimensional thermodynamic limit, we need an extrapolation of the system size with a fixed aspect ratio as in Ref \cite{White2007}.
It was known that the finite size effect varies with different aspect ratios and there exists an optimal aspect ratio with which the finite size effect is minimum \cite{White2007}. 
In this work, we also study systems with different aspect ratios trying to find the optimal value. Because pinning fields are applied in the open edges of the studied systems, we define the aspect ratio $\alpha = (L_\mathrm{x}-2) / L_\mathrm{y}$ \cite{White2007}.

It was shown in Ref \cite{White2007} that in the DMRG calculation, the local physical quantities also scale linearly with the truncation error, similar to the ground state energy. In Fig.~\ref{Fit_4/3}, we illustrate the variation of magnetization (defined in Eq.~(\ref{eq:m})) as a function of the truncation error $\varepsilon$ for $\alpha = 4/3$ and $L_\mathrm{y} = 6$, $9$, $12$.
From Fig.~\ref{Fit_4/3}, we can see a linear dependence of $M$ on $\varepsilon$, which allows us to perform reliable extrapolation to obtain the zero truncation error value. We notice that the error bar for the $L_y = 6$ system results from the derivation of the magnetization of the three sublattices to the ratio $-1: 2: -1$, while the error from finite bond dimension in DMRG is small for large bond dimensions. The results for other ratios ($3/3$, $5/3$, $6/3$) are shown in Appendix.~\ref{sec:appendix A}.

\section{Results}
\label{sec:III}

\subsection{Magnetization}

In Ref \cite{White2007}, it is found that there exists a critical aspect ration $\alpha_c$ at which the magnetization remains almost constant with system sizes. Finite size results for systems with aspect ratio around $\alpha_c$ allow us to obtain accurate values for magnetization in the two-dimensional thermodynamic limit.

\begin{figure}[h]
	\includegraphics[width=85mm]{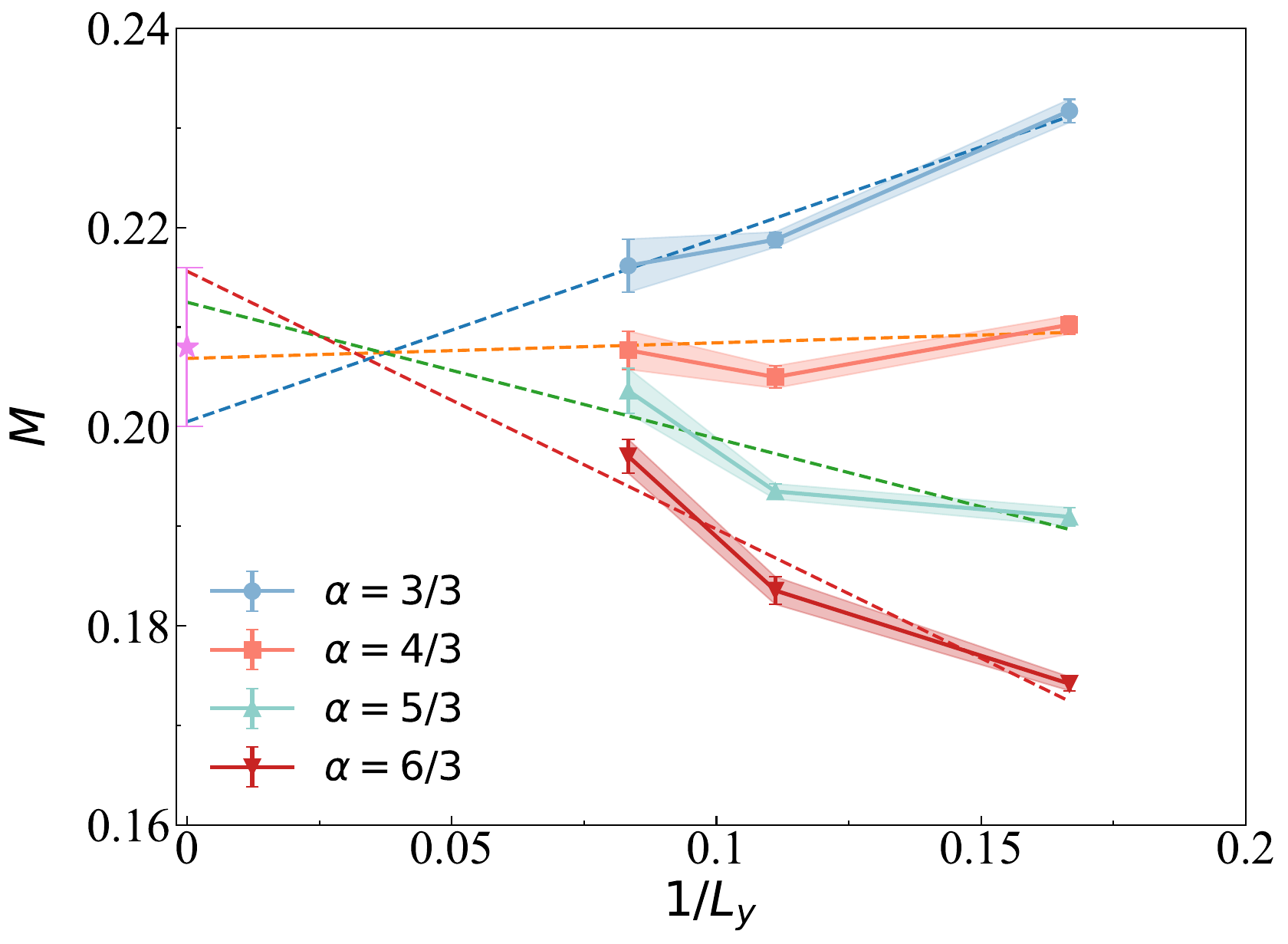}
	\caption{DMRG results for magnetization $M$ of TLHM as a function of $L_\mathrm{y}$ for different aspect ratios $\alpha$ with $L_\mathrm{y} = 6$, $9$, and $12$. The dashed lines represent the linear extrapolations for $\alpha=3/3$, $4/3$, $5/3$, and $6/3$ respectively. Based on these results, the critical aspect ratio is estimated to be around $\alpha_c = 4/3$ and a conservative estimation of magnetization in the thermodynamic limit is $M_0 = 0.208(8)$ (the pink star).}
	\label{Mc_Ly}
\end{figure}

In the vicinity of $\alpha_c$, the magnetization is a function of both aspect ratio $\alpha$ and the width of the system $L_\mathrm{y}$ with the following relationship \cite{White2007},
\begin{equation}
M (\alpha,L_\mathrm{y})  = M_0  + \beta(\alpha - \alpha_c)/L_\mathrm{y}
\label{eq:fit}
\end{equation}
where $\beta$ is a negative parameter independent of $\alpha$ \cite{foot1}.

In previous study \cite{White2007}, the value $\alpha_c$ for TLHM was determined to be $1.6\sim 1.7$. Based on this, we study systems with $\alpha$ = $3/3$, $4/3$, $5/3$, and $6/3$. The variation of $M$ for these aspect ratios with system size $1/L_\mathrm{y}$ are shown in Fig.~\ref{Mc_Ly}. Even though we only present results for width $L_y = 6, 9,$ and $12$ due to the constraint from boundary conditions, we find that results with $\alpha=4/3$ have the smallest finite size effect and are likely to be close to $\alpha_c$ as shown in Fig.~\ref{Mc_Ly},  We attribute the discrepancy of $\alpha_c$ between our result and the value in Ref \cite{White2007} to the small system size considered in Ref \cite{White2007}.

\begin{figure}[t]
	\includegraphics[width=85mm]{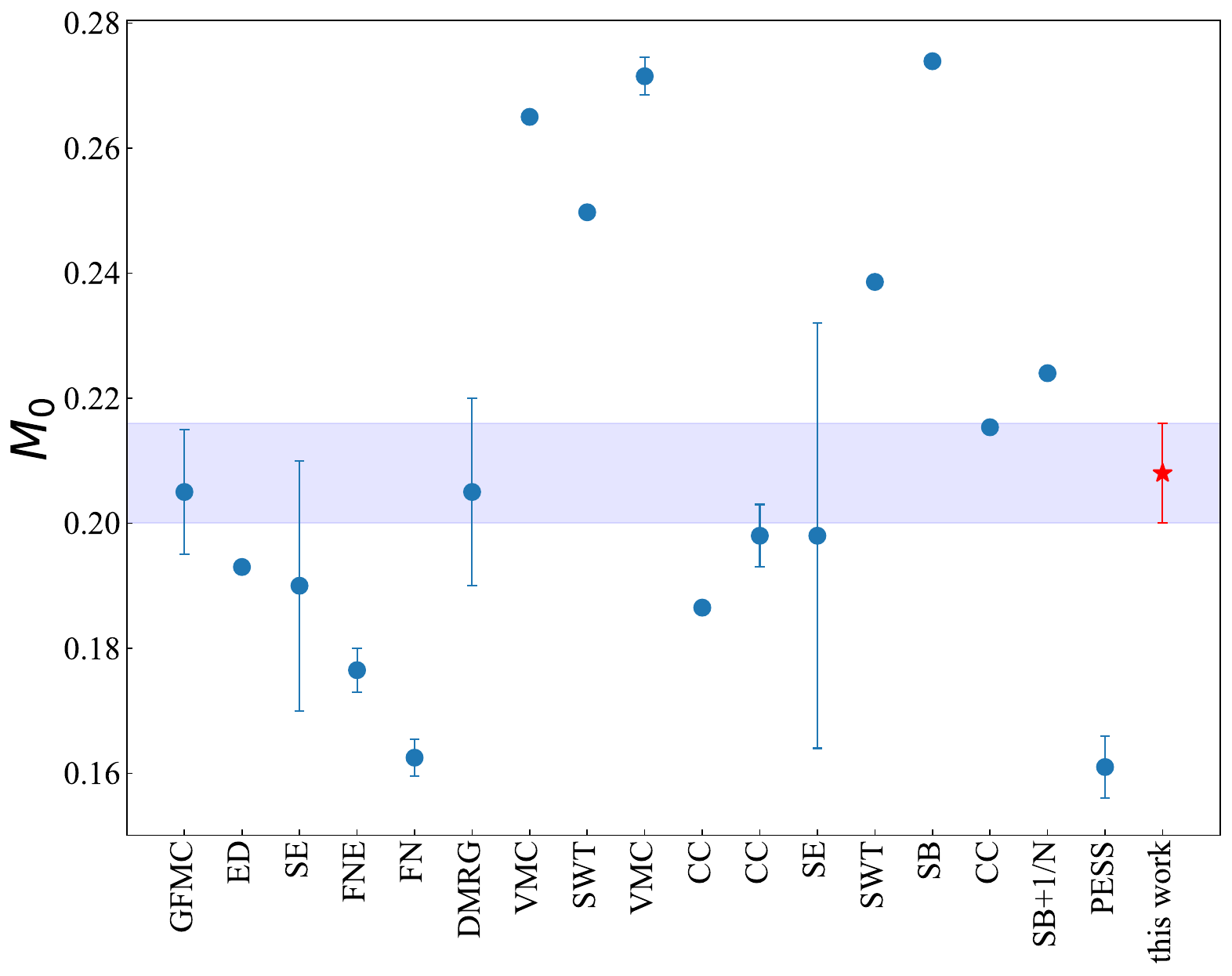}
	\caption{
		Comparison of the results of $M_0$ for TLHM obtained by different methods. These data were obtained from GFMC \cite{GFMC}, ED \cite{ED2004}, SE \cite{SE2006,SB2015}, FNE \cite{FN2006}, FN \cite{FN2006}, DMRG \cite{White2007}, VMC \cite{VMC2009,VMC2014}, SWT \cite{SB2015}, CC \cite{CC2014,CC2015,CC2016}, SB \cite{SB2015,SB+1/N}, and PESS \cite{PESS2022}.}
	\label{M_c_all}
\end{figure}

From Eq.~(\ref{eq:fit}), we know that when $\alpha$ is less (larger) than $\alpha_c$, $M$ decreases (increases) with the increase of $L_\mathrm{y}$. In Fig.~\ref{Mc_Ly}, we find the magnetization for $\alpha=3/3$ decreases with $1/L_\mathrm{y}$ approaching zero, while the results for $\alpha=5/3$ and $6/3$ both increases with $1/L_\mathrm{y}$ approaching zero, consistent with the
relationship in Eq.~(\ref{eq:fit}). We also find that magnetization of the three aspect ratios near $\alpha=4/3$ converges towards the result of $\alpha=4/3$. With a simple extrapolation of the results, a conservative estimation of the magnetization in the thermodynamic limit ($M_0$) can be obtained as $0.208(8)$.

We also show a comparison of our results with the previous results in the literature in Fig.~\ref{M_c_all}. It is noteworthy that the results obtained through Green's Function Monte Carlo (GFMC) \cite{GFMC}, Series Expansion (SE) \cite{SE1993,SE2006}, Coupled Cluste (CC) \cite{CC2015,CC2016}, and previous DMRG works \cite{White2007} closely align with our result.

\subsection{Ground state energy}

\begin{figure}[t]
	\includegraphics[width=85mm]{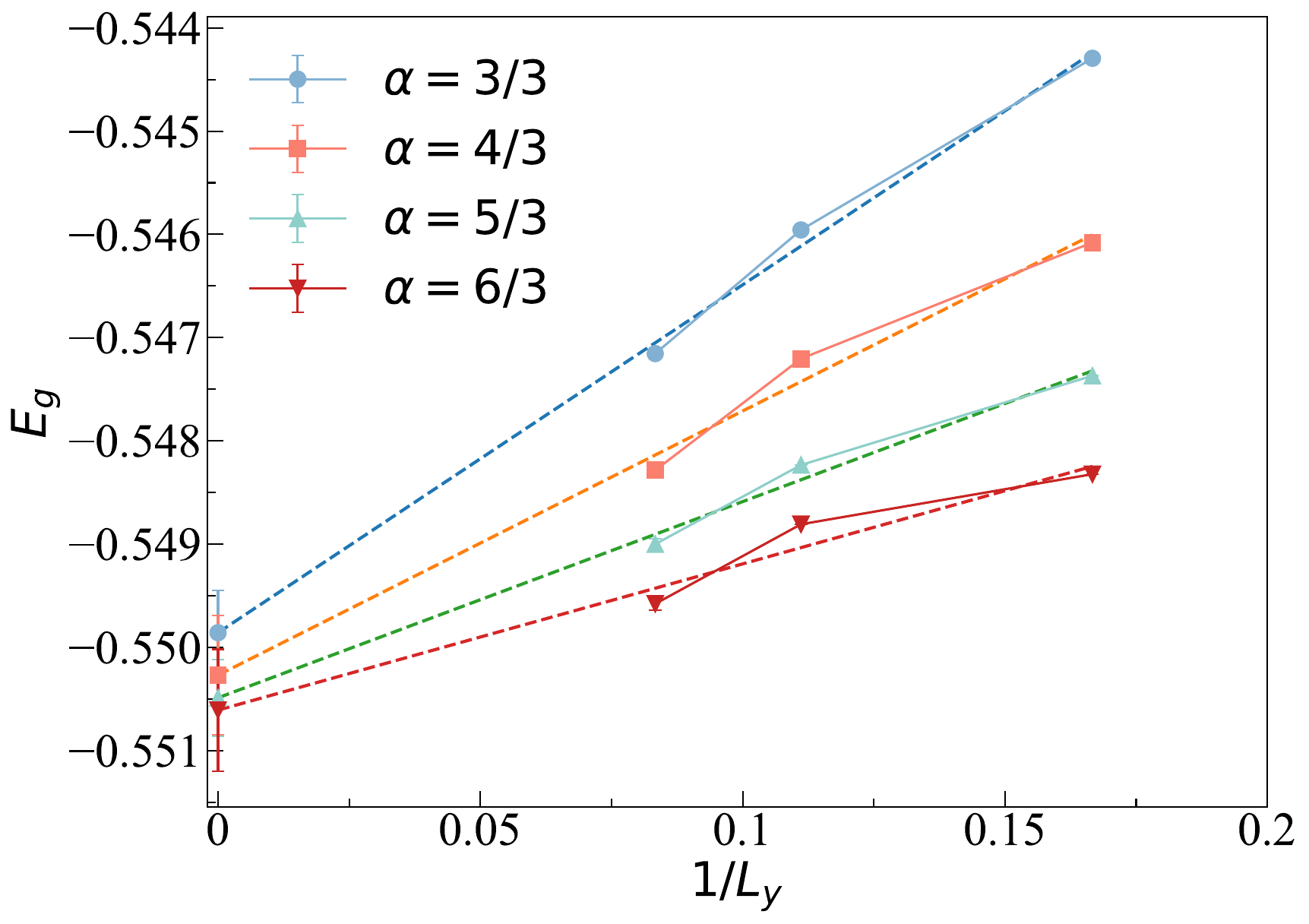}
	\caption{The ground state energy per site with different aspect ratios $\alpha$ as a function of $L_\mathrm{y}$. All data points have been linearly extrapolated to truncation error $\varepsilon = 0$. The dashed lines represent the linear fits for different $\alpha$. We can find that the energy with different $\alpha$ are extrapolated to the same value with small uncertainty. From these results, the ground state energy per site in the thermodynamic limit is estimated to be $E_g = -0.5503(8)$.}
	\label{energy}
\end{figure}

We show the ground state energy per site $E_g$ as a function of $1/L_\mathrm{y}$ in Fig.~\ref{energy} for $\alpha$ ranging from $3/3$ to $6/3$. The values in  Fig.~\ref{energy} are obtained from the extrapolation with truncation errors in DMRG calculation. In Fig.~\ref{energy}, we can find that the linear extrapolations of the energies for different $\alpha$ converge to the same value with small uncertainty. The estimated ground state energy per site is $E_g = -0.5503(8)$ which is consistent with recent calculations with Projected Entangled Simplex State (PESS) \cite{PESS2022}, CC \cite{CC2016}, SE \cite{SE2006} and DMRG \cite{DMRG2016}.

\section{Conclusion}
\label{sec:IV}
In conclusion, we provide an accurate value for the magnetization of the $120^\circ$ order in the spin-1/2 antiferromagnetic Heisenberg model on the triangular lattice.
With careful extrapolation with truncation error and suitable finite size scaling, we give a conservative estimation of the magnetization as
$M_0 = 0.208(8)$. We also find that the critical aspect ratio $\alpha_c$ is approximately $4/3$. The ground state energy per site we obtain is $E_g = -0.5503(8)$,
agreeing well with previous results. The accurate magnetization and energy provide useful benchmark values for both calibrations of existing methods and the development of new methods in the future.

\begin{acknowledgments}
We thank useful discussion with Tao Xiang, Haijun Liao and Zhiyuan Xie.
The calculation in this work is carried out with TensorKit \cite{foot2}.
The computations in this paper were run on the Siyuan-1 cluster supported by the Center for High Performance Computing at Shanghai Jiao Tong University.
MQ acknowledges the support from the National Key Research and Development Program of MOST of China (2022YFA1405400), the Innovation Program for Quantum Science and Technology (2021ZD0301902),
the National Natural Science Foundation of China (Grant
No. 12274290) and the sponsorship from Yangyang Development Fund.

\end{acknowledgments}

\bibliography{reference}

\begin{thebibliography}{46}
\bibitem[{foo({\natexlab{a}})}]{foot1}
\bibinfo{note}{{From simple linear extrapolations of the results in Fig.~3, we
  find the slope depend linearly on $\alpha$}}.
\end{thebibliography}

\appendix

\section{extrapolation with truncation error}
\label{sec:appendix A}

\begin{figure*}[t]
	\centering

	\includegraphics[height=45mm]{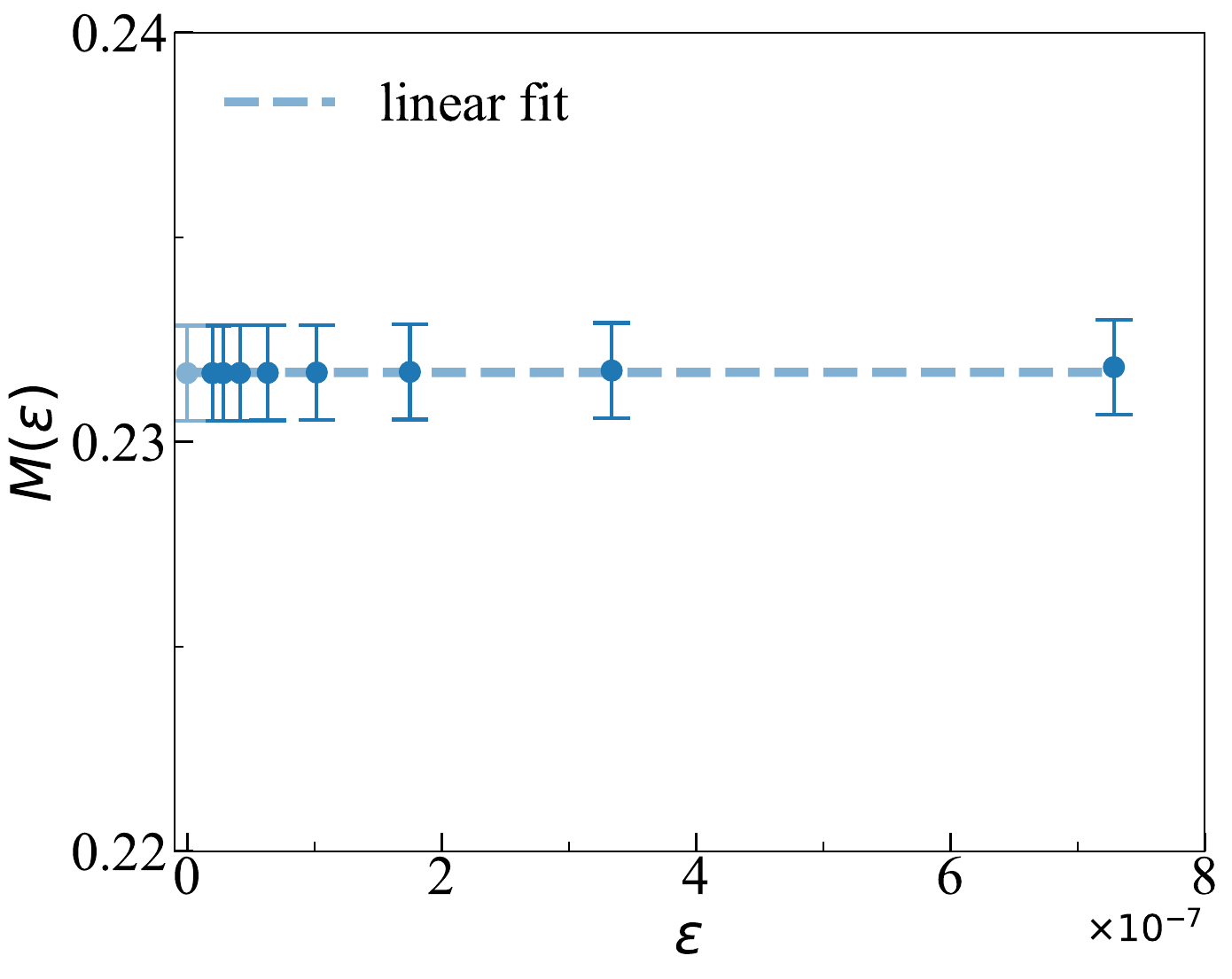}
	\includegraphics[height=45mm]{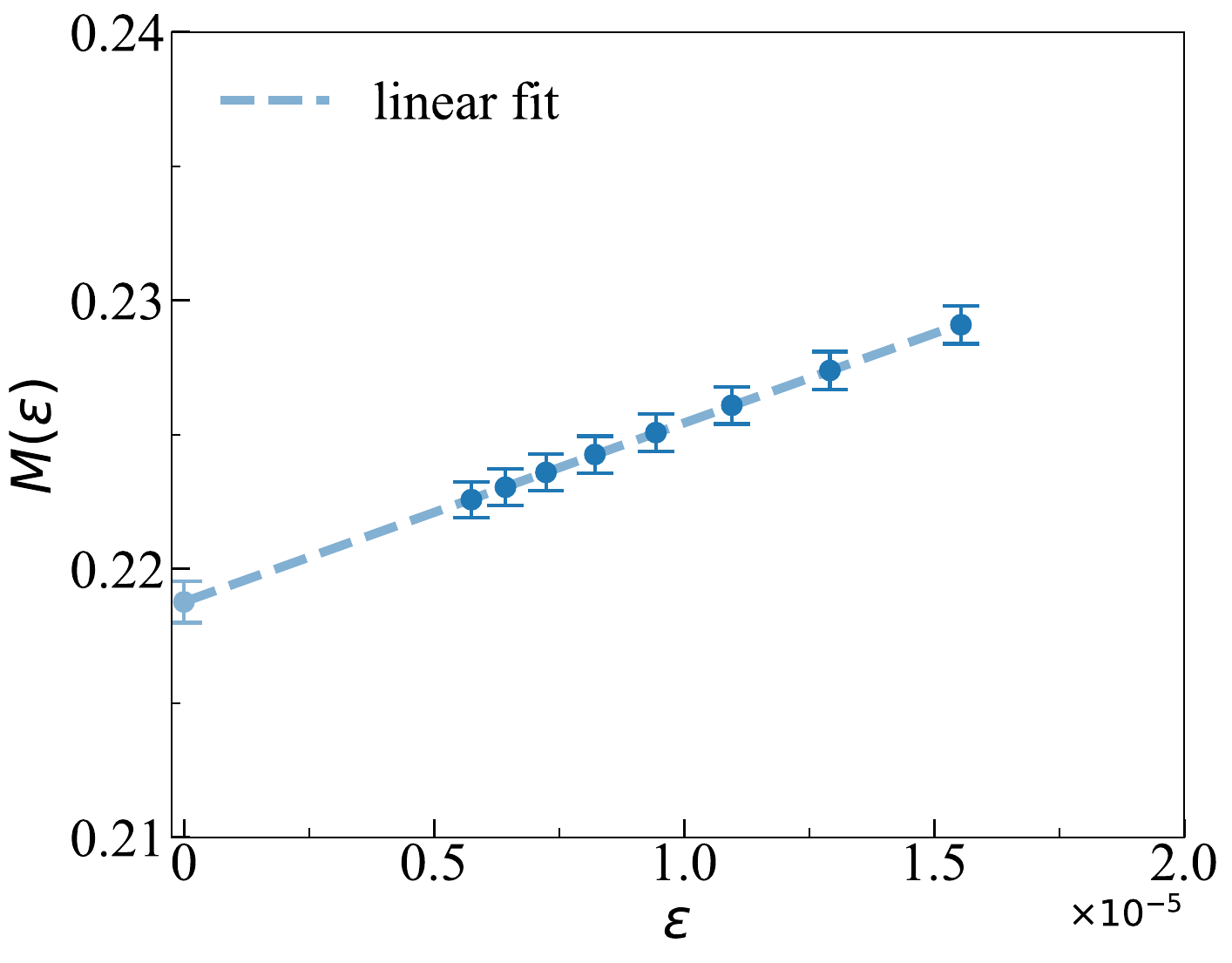}
	\includegraphics[height=45mm]{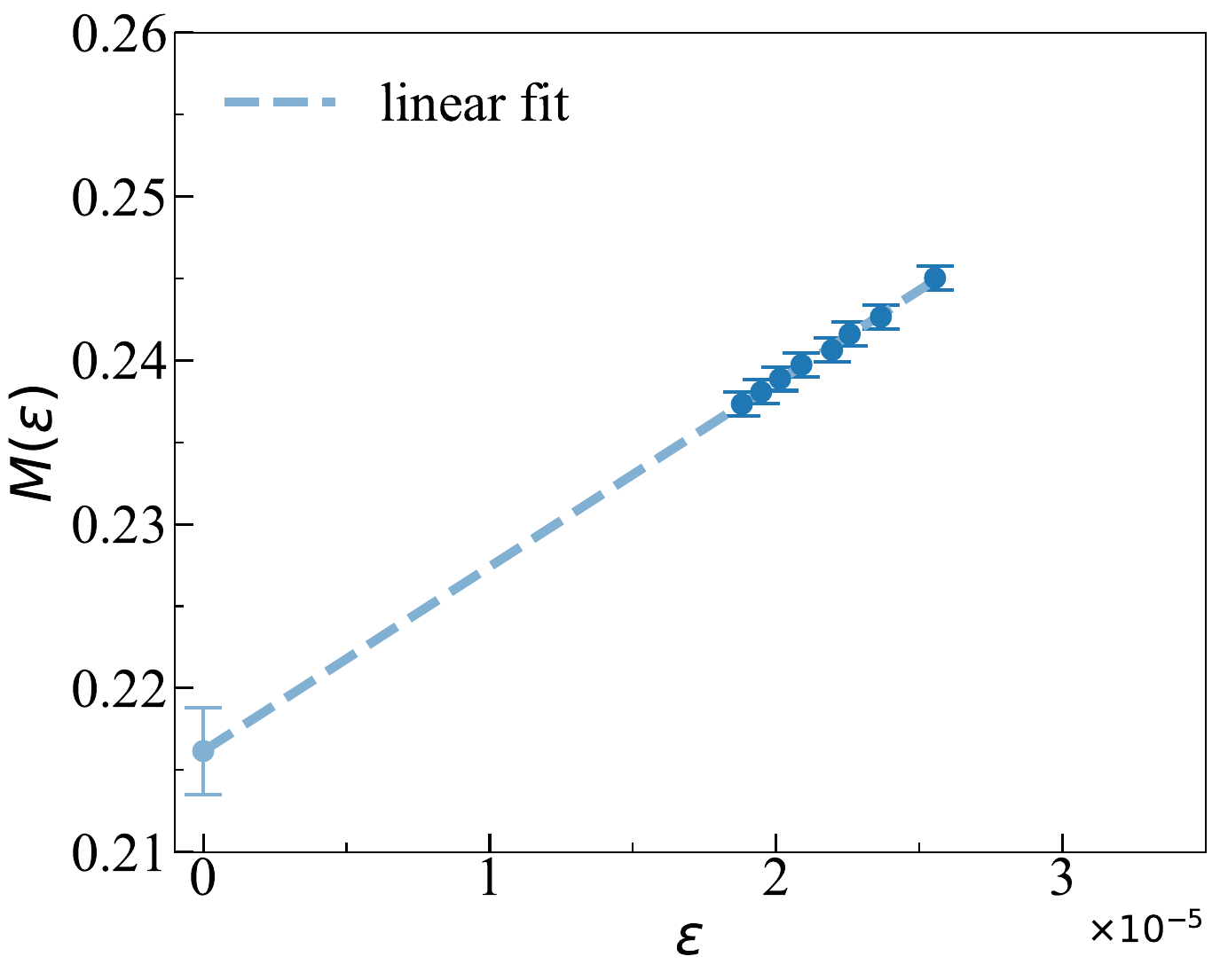}

	\includegraphics[height=45mm]{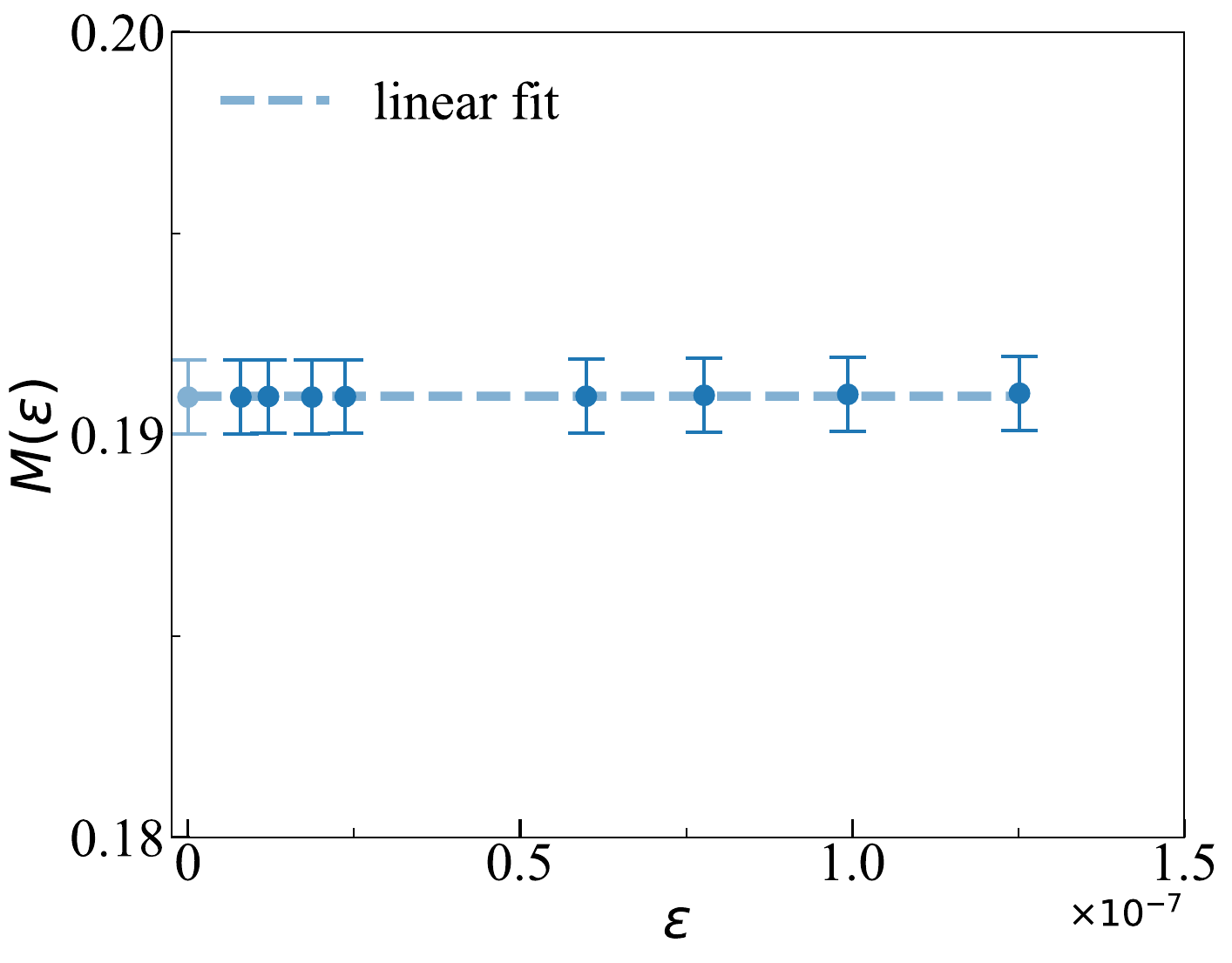}
	\includegraphics[height=45mm]{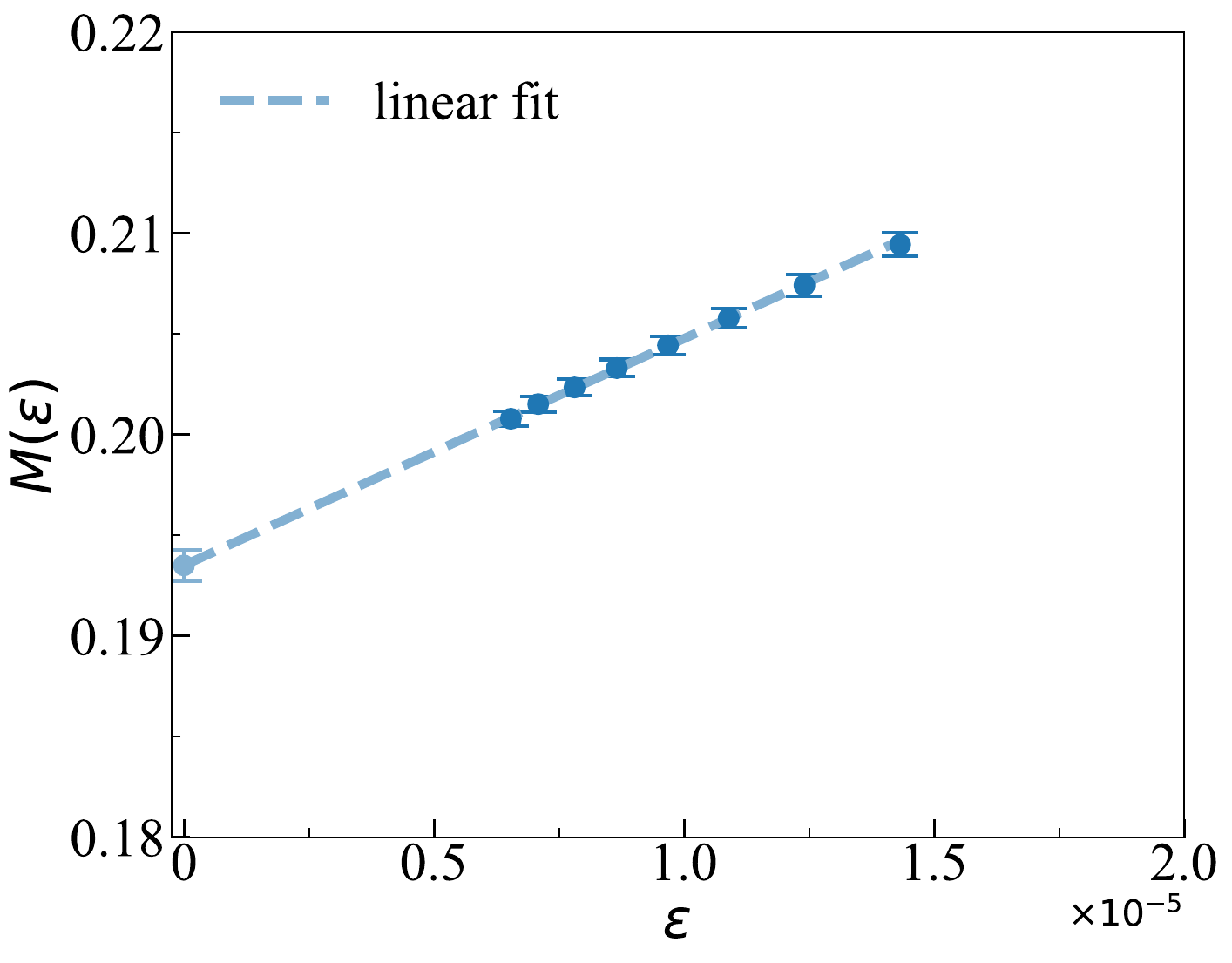}
	\includegraphics[height=45mm]{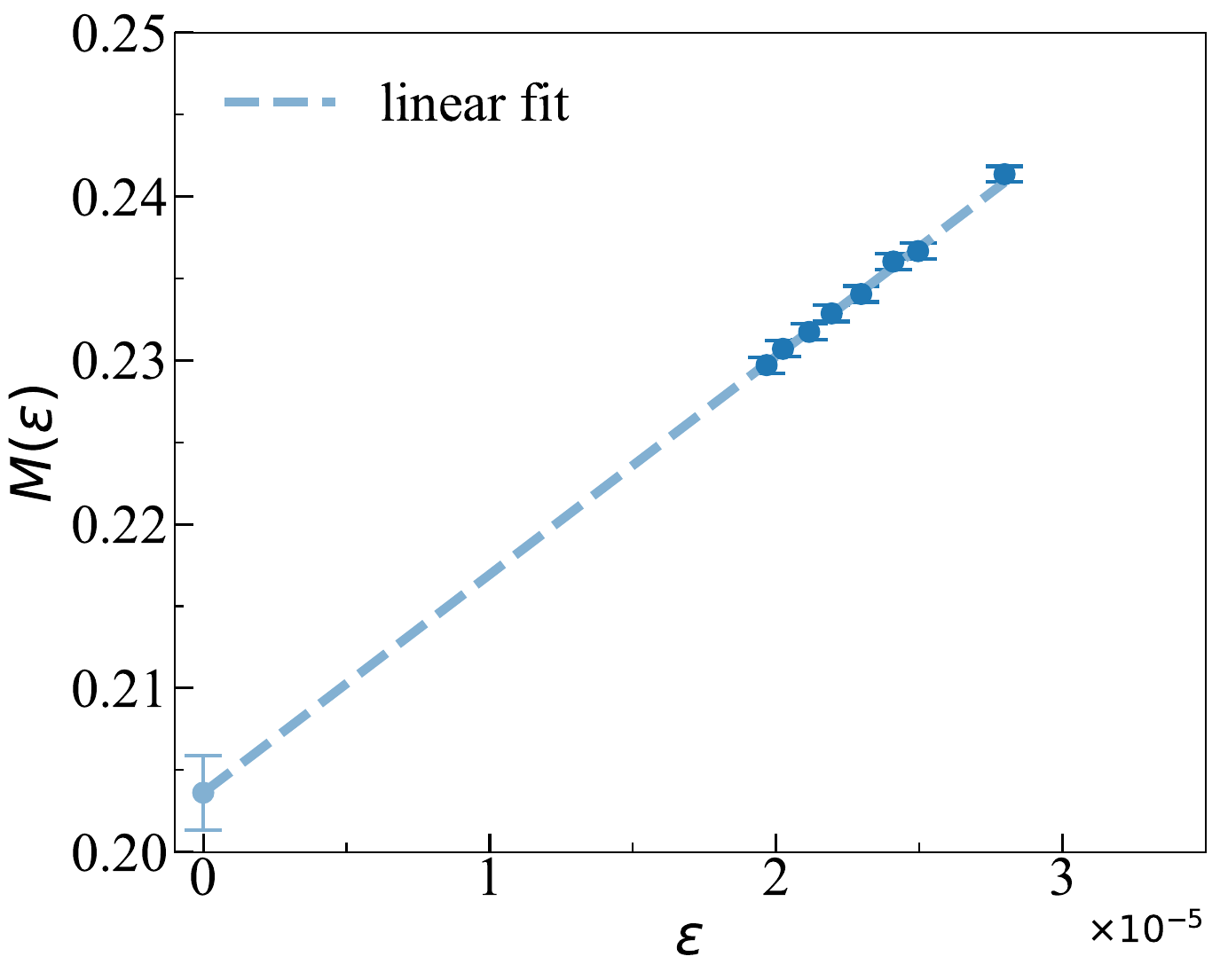}

	\includegraphics[height=45mm]{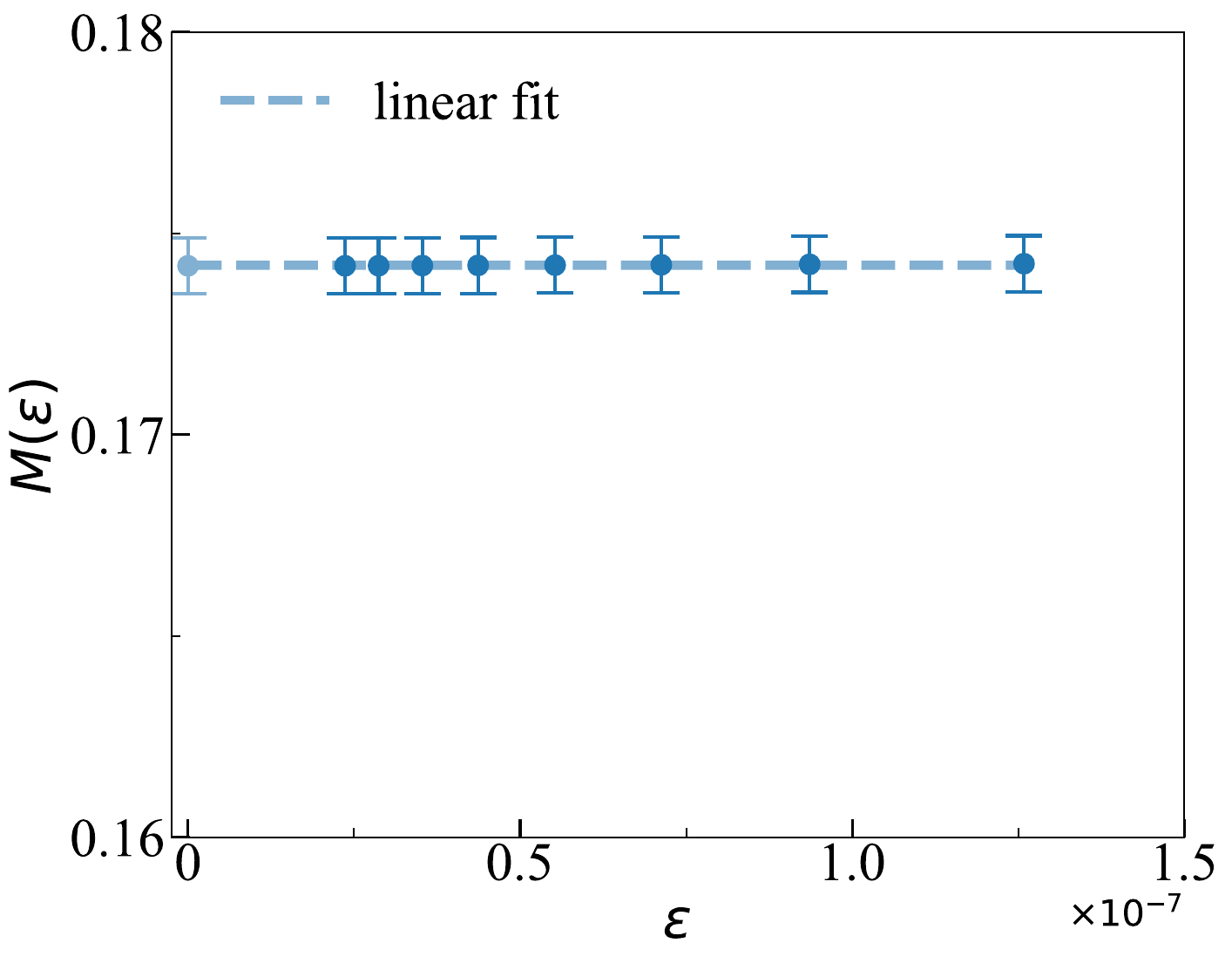}
	\includegraphics[height=45mm]{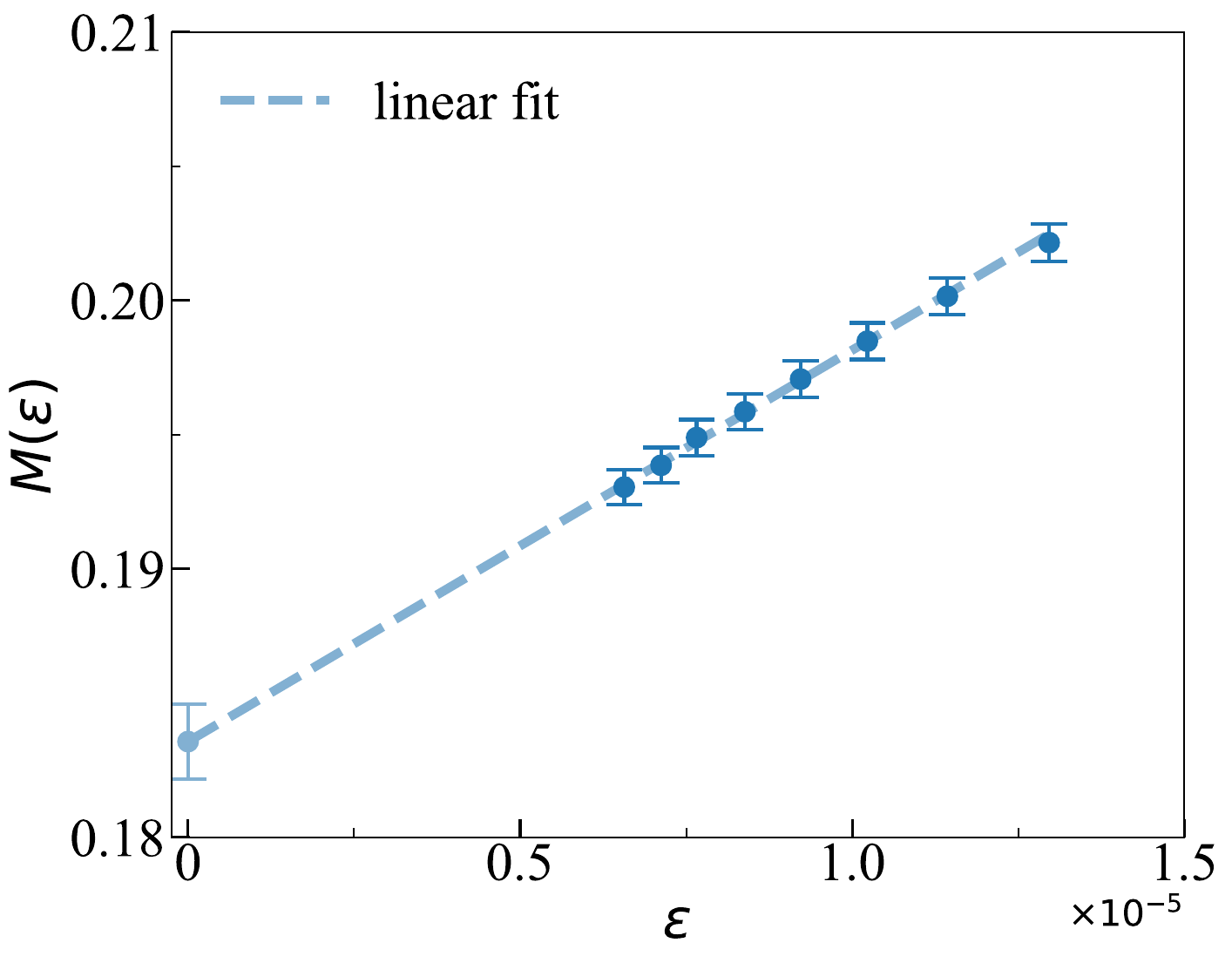}
	\includegraphics[height=45mm]{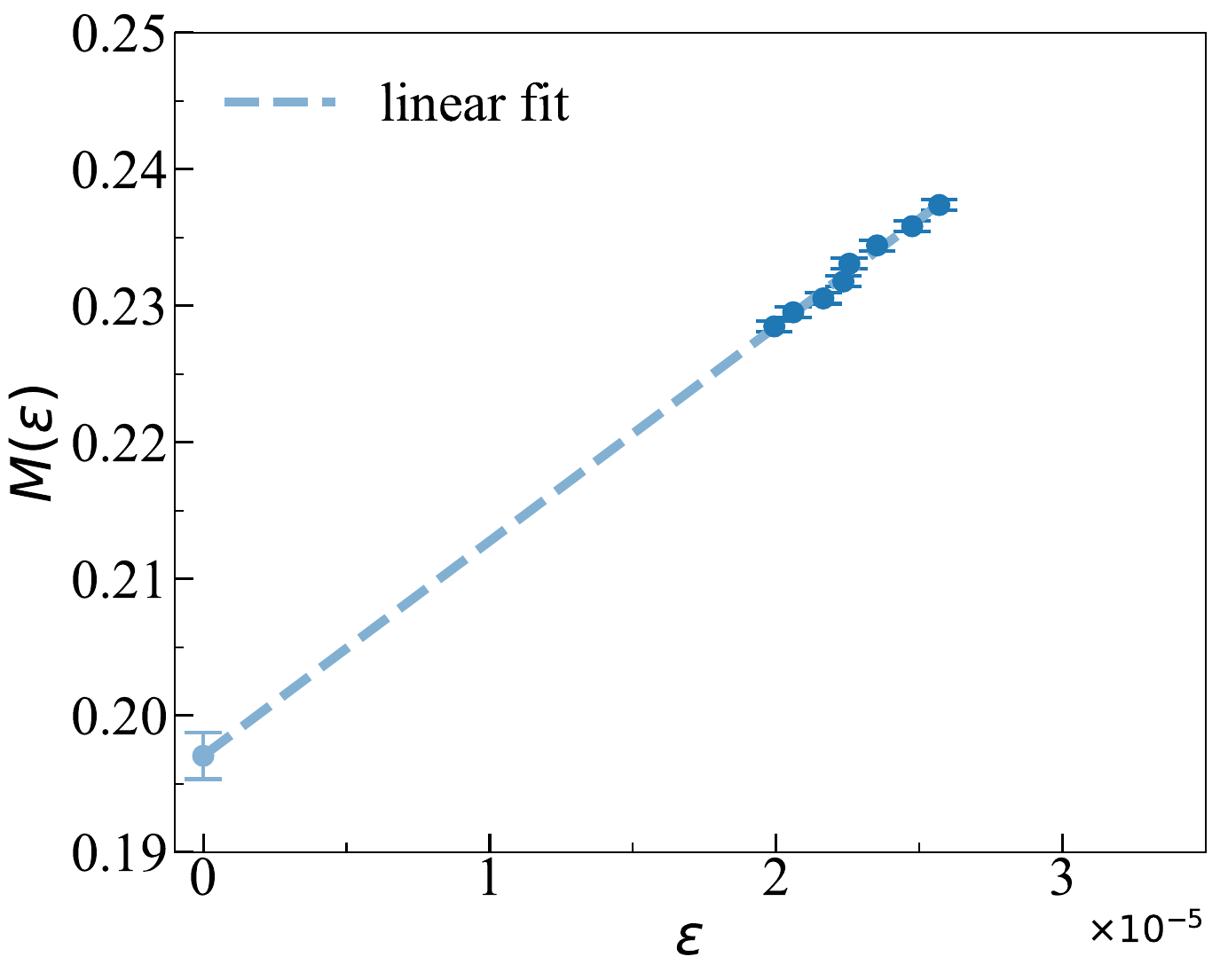}

	\caption{From top to bottom, the extrapolation of magnetization with truncation error in DMRG calculation for aspect ratio $\alpha = 3/3$, $5/3$, and $6/3$. The result for $L_\mathrm{y} = 6$, $9$, and $12$ are displayed from left to right. The dashed lines represent linear fits.}
	\label{Fit_4/3_comp}

\end{figure*}

In Fig.~\ref{Fit_4/3_comp}, similar as in Fig.~\ref{Fit_4/3}, we show the extrapolation of magnetization with truncation error in DMRG calculation for aspect ratio $\alpha = 3/3$, $5/3$, and $6/3$ with width $L_\mathrm{y} = 6$, $9$, and $12$. 

\end{document}